\documentclass[twocolumn,prd,showpacs,preprintnumbers,
               superscriptaddress,floatfix]{revtex4}

\usepackage{bm}

\begin{document}

\title{S wave meson spectra from the light cone harmonic oscillator
       model with a consistent hyperfine interaction}

\author{Shan-Gui Zhou}
 \email{sgzhou@mpi-hd.mpg.de}
 \affiliation{Max-Planck-Institut f\"ur Kernphysik, D-69029
              Heidelberg, Germany}
 \affiliation{School of Physics, Peking University, Beijing 100871, China}
 \affiliation{Institute of Theoretical Physics, Chinese Academy of Sciences,
              Beijing 100080, China}
 \affiliation{Center of Theoretical Nuclear Physics, National Laboratory of
              Heavy Ion Accelerator, Lanzhou 730000, China}
\author{Hans-Christian Pauli}
 \email{pauli@mpi-hd.mpg.de}
 \affiliation{Max-Planck-Institut f\"ur Kernphysik, D-69029
              Heidelberg, Germany}

\date{November 21, 2003}

\begin{abstract}
We use a light cone harmonic oscillator model to study S wave
meson spectra, namely the pseudoscalar and vector mesons. The
model Hamiltonian is a mass squared operator consisting of a
central potential (a harmonic oscillator potential) from which a
hyperfine interaction is derived. The hyperfine interaction is
responsible for the splitting in the pseudoscalar-vector spectra.
With 4 parameters for the masses of up/down, strange, charm and
bottom quarks, 2 for the harmonic oscillator potential and 1 for
the hyperfine interaction, the model presents a reasonably good
agreement with the data.
\end{abstract}


\maketitle

\section{Introduction}

The effective light cone Hamiltonian, a mass squared operator,
consists of a central potential and corresponding fine and
hyperfine interactions~\cite{BroPauPin98}. With the central (and
the confinement) potential approximated by a harmonic oscillator
potential and a hyperfine interaction, a Dirac delta interaction,
assumed to act on the vector meson only, the model gave an
universal description of S wave pseudoscalar and vector meson
spectra~\cite{FrePauZho02a,FrePauZho02b}. By introducing a
phenomenological spin orbit interaction with one additional
parameter, this model was also applied to study P wave $D_s$
mesons~\cite{ZhoPau03,Zho03}.

In the model proposed in~\cite{FrePauZho02a,FrePauZho02b}, the
hyperfine interaction, the Dirac delta interaction, must be
renormalized. The renormalization parameter, in fact, is hidden in
the renormalization procedure, thus depending on the pseudoscalar
sector. In other words, for each sector of pseudoscalar mesons,
one has an additional parameter for the renormalization. The
situation is certainly not satisfactory.

One knows that the hyperfine interaction is in
general~\cite{BjoDre64},
\begin{equation}
 \textstyle
 V_\mathrm{hf} = \frac{\bm\nabla^2V(r)}{6m_1m_2}
                 \bm\sigma_1\bm\sigma_2 ,
\end{equation}
where $V(r)$ is the central potential in the model Hamiltonian. If
$V(r)$ is the Coulomb potential, one gets for the hyperfine the
Dirac delta interaction~\cite{BjoDre64}. If $V(r)$ is a harmonic
oscillator potential,
\begin{equation}
 \textstyle
 V(r) = -a + \frac{1}{2} f r^2 ,
 \label{eq:ho}
\end{equation}
however, one gets a much simpler hyperfine interaction,
\begin{equation}
 \textstyle
 V_\mathrm{hf} = \frac{f}{2m_1m_2}
                 \bm\sigma_1\bm\sigma_2 , \
 \bm\sigma_1\bm\sigma_2
 = \left\{\matrix{-3, & \mathrm{for}\ S=0,\cr
                  +1, & \mathrm{for}\ S=1.}\right.
 \label{eq:hf}
\end{equation}
The hyperfine interaction acts on both triplet and singlet states.

In the present work, we will use the harmonic oscillator potential
as the central potential and the confinement. This potential gives
naturally the Anisovich law~\cite{AniAniSar00} as shown in
\cite{FrePauZho02a,FrePauZho02b}. For the hyperfine interaction,
we adopt the more general one as given in (\ref{eq:hf}). In
section~\ref{sec:para} the model parameters are determined. The
results is given in section~\ref{sec:results}. Finally a summary
is presented in section~\ref{sec:summary}.

\section{Parameter determination}
\label{sec:para}

An effective light cone Hamiltonian which is a mass square
operator was given in~\cite{Pau00b,Pau03a},
\begin{eqnarray}
   H_\mathrm{eHC} \Psi = M^2 \Psi
\,,
\end{eqnarray}
where $M^2$ the mass squared of the meson in question and
\begin{eqnarray}
 H_\mathrm{eLC}
 & = & \textstyle
 \left(m_1+m_2\right)^2 + 2\left(m_1+m_2\right) H ,
 \\
 H & = & \textstyle
 \left( -\frac{\bm \nabla^2}{2m_r} + V(r) + V_\mathrm{hf} \right) ,
 \\
 M^2
 & = &
 \left(m_1+m_2\right)^2 + 2\left(m_1+m_2\right) E .
 \label{eq:M2E}
\end{eqnarray}
$m_1$ and $m_2$ are the constituent quark masses and $E$ the
eigenvalue of $H$
\begin{equation}
 \textstyle
 E = -a + (2n+\frac{3}{2}) \omega ,
 \label{eq:E}
\end{equation}
with $\omega\equiv \sqrt{f/m_r}$ and $m_r=m_1m_2/(m_1+m_2)$.

In order to describe S wave mesons except for those consisting of
at least one top quark, we need six parameters if we assume
$m_u=m_d$. The parameters include: four masses for the up/down,
strange, charm and bottom quarks; the depth $a$ and the spring
constant $f$ for the harmonic oscillator potential. One can first
determine $m\equiv m_{u/d}$, $a$ and $f$ by the following
experimentally well determined masses,
\begin{eqnarray}
 \begin{array}{rcl}
  M^2_{\rho^+(2S)}
  & = &
  4m^2 + 2m\left(-2 a+3\omega+\textstyle \frac{f}{m^2}\right)
  + 8m\omega ,
  \\
  M^2_{\rho^+(1S)}
  & = &
  4m^2 + 2m\left(-2 a+3\omega+\textstyle \frac{f}{m^2}\right) ,
  \\
  M^2_{\pi^+(1S)}
  & = &
  4m^2 + 2m\left(-2 a+3\omega-\textstyle \frac{3f}{m^2}\right) .
 \end{array}
\end{eqnarray}
Three differences are available from the experiment:
\begin{eqnarray}
 \begin{array}{rcl}
  \Delta^2
  & = & \phantom{-} M^2_{\rho^+(2S)} - M^2_{\rho^+(1S)} ,
  \\
  D^2
  & = & \phantom{-} M^2_{\rho^+(1S)} - M^2_{\pi^+(1S)} ,
  \\
  A^2 & = &  - M^2_{\rho^+(1S)} + \frac34\Delta^2 + \frac14 D^2 .
 \end{array}
\end{eqnarray}
In terms of the model parameters they are:
\begin{equation}
 \textstyle
 \Delta^2 = 8 m\omega ,\
 D^2 = \frac{8 f}{ m} ,\
 A^2 = 4 a m - 4 m^2 .
\end{equation}
One then has
\begin{equation}
 \textstyle
 m = \frac{\Delta^2}{4 D} ,\
 a = \frac{A^2 D}{\Delta^2} + \frac{\Delta^2}{4 D} ,\
 f = \frac{\Delta^2 D}{32} .
\end{equation}
The mass for the quark $q$ ($q$ = $s$, $c$ and $b$) is determined
from (\ref{eq:M2E}) and (\ref{eq:E}) with $M$ the mass of the
vector meson consisting of $u$ and $q$.

\begin{table}
\caption{\label{table:dataforpara} Experimental masses (in GeV) of
mesons used to determine the parameters of the light cone harmonic
oscillator model.}
%\begin{ruledtabular}
\begin{tabular}{||@{$\;\:\,$}c@{$\;\:\,$}|@{$\;\:\,$}c@{$\;\:\,$}
                 |@{$\;\:\,$}c@{$\;\:\,$}|@{$\;\:\,$}c@{$\;\:\,$}
                ||@{$\;\:\,$}c@{$\;\:\,$}|@{$\;\:\,$}c@{$\;\:\,$}||}
 \hline\hline
 ${\pi^+}$  & ${\rho^+}$       & ${\rho^+(1450)}$  &
 ${K^{*+}}$ & ${\bar D^{*0}}$  & ${B^{*+}}$        \\
 \hline\hline
 0.1396     & 0.7711           & 1.465             &
 0.8917     & 2.0067           & 5.325             \\
 \hline\hline
\end{tabular}
%\end{ruledtabular}
\end{table}

\begin{table}
\caption{The parameters of the light cone harmonic oscillator
model. Masses and $a$ in GeV. $f$ in GeV$^3$.
%Parameters given by Frederico, Pauli and Zhou~\protect\cite{FrePauZho02b}
%are included for comparison.
}~\label{table:para}
%\begin{ruledtabular}
\begin{tabular}{||@{$\;\;\;$}c@{$\;\;\,$}|@{$\;\;\,$}c@{$\;\;\,$}
                 |@{$\;\;\,$}c@{$\;\;\,$}|@{$\;\;\,$}c@{$\;\;\,$}
                 |@{$\;\;\,$}c@{$\;\;\,$}|@{$\;\;\,$}c@{$\;\;\;$}||}
 \hline\hline
 $m_{u/d}$ & $m_s$  & $m_c$  & $m_b$  & $a$    & $f$          \\
 \hline\hline
 \multicolumn{6}{||c||}{Without the fudge factor}  \\
 \hline
 0.5115    & 0.7035 & 1.9211 & 5.2628 & 0.8599 & 0.0368       \\
 \hline\hline
 \multicolumn{6}{||c||}{The fudge factor $f^*=0.3$}\\
 \hline
 0.2802    & 0.5685 & 1.8380 & 5.2027 & 0.9163 & 0.0671       \\
 \hline\hline
\end{tabular}
%\end{ruledtabular}
\end{table}

In Table~\ref{table:dataforpara} are given the experimental meson
masses which are used for fixing the model parameters. It turns
out that the parameters thus determined are not reasonable as seen
in Table~\ref{table:para} although those parameters do produce
good agreement with the data. Particularly, $m_{u/d}$ and $m_s$
are too large. The reason for the large $m_{u/d}$ and $m_s$ is
that the hyperfine interaction (\ref{eq:hf}) which is fixed
completely by $V(r)$ is so large. In order to get a reasonable set
of constituent quark masses, we introduce a parameter which
reduces the hyperfine interaction on purpose. With this additional
parameter called the fudge factor, $f^*$, the hyperfine
interaction reads,
\begin{equation}
 \textstyle
 V_\mathrm{hf} =  f^* \frac{f}{2m_1m_2} \bm\sigma_1\bm\sigma_2 .
 \label{eq:hf-suppressed}
\end{equation}
We found that $f^*=0.3$ produces reasonable values for the model
parameters as given in Table~\ref{table:para}.

We note that the fudge factor $f^*$ accounts for shortcomings of
our certainly oversimplified model. This factor is universal in
the sense that once fixed, it is valid for all S wave mesons.
Furthermore, even with this fudge factor, the number of parameters
of the present model is still less than many other models proposed
for the description of meson spectra.

\section{Results and discussion}
\label{sec:results}

The (flavor off-diagonal) S wave meson spectra calculated from the
present model are given in Tables~\ref{table:ud}, \ref{table:us}
and \ref{table:heavy}. Since one can quite easily calculate these
meson spectra from Eqs.~(\ref{eq:M2E}) and (\ref{eq:E}) with
parameters given in Table~\ref{table:para}, we do not list
theoretical masses for those excitations which are not observed
yet. For comparisons, the available data from Hagiwara \textit{et
al}~\cite{RPP02} and other theoretical masses are included in
these tables. The agreement between the admittedly simple model
and the experiment is generally very good. Particularly, the
present model reproduces well the masses for heavier mesons.

\begin{table}
\caption{\label{table:ud} S wave spectra for light unflavored
mesons. Masses in GeV. The pion mass
$m_{\pi^+}=139.57018\pm0.00035$ MeV is accurately known, but only
the first 4 digits are used here.}
%\begin{ruledtabular}
\begin{tabular}{||l|l|l@{$\;\;\;\;\;\;\;\;$}||l|l|l@{$\;\;\;\;\;\;\;\;$}||}
 \hline\hline
 $n$ & Experiment$^1$ & Theory    & $n$ & Experiment$^1$& Theory    \\
 \hline\hline
 \multicolumn{3}{||c||}{$^1\mathrm{S}_0$ Singlets $\pi^+$} &
 \multicolumn{3}{  c||}{$^3\mathrm{S}_1$ Triplets $\rho^+$} \\
 \hline
 1   & 0.1396(0)      & 0.1396    & 1 & 0.7711(9)     & 0.7711    \\
     &                & 0.140$^2$ &   &               & 0.768$^2$ \\
     &                & 0.15$^3$  &   &               & 0.77$^3$  \\
 2   & 1.300(100)     & 1.2650    & 2 & 1.465(25)     & 1.4650    \\
     &                & 1.223$^2$ &   &              & 1.465$^2$ \\
     &                & 1.30$^3$  &   &               & 1.45$^3$  \\
 3   & 1.801(13)      & 1.7950    & 3 & 1.700(20)$^a$ & 1.9230    \\
     &                & 1.739$^2$ &   &               & 1.924$^2$ \\
     &                & 1.88$^3$  &   &               & 2.00$^3$  \\
 4   & ---            & 2.1620    & 4 & 2.150(17)     & 2.2912    \\
     &                & 2.317$^2$ &   &               & 2.292$^2$ \\
 \hline\hline
 \multicolumn{6}{l}{$^1$Hagiwara \textit{et al}~\cite{RPP02};
                    $^2$Frederico, Pauli and Zhou~\cite{FrePauZho02b};} \\
 \multicolumn{6}{l}{$^3$Godfrey and Isgur~\cite{GodIsg85}. % } \\
                    $^a$Could be a $D$-wave state~\cite{AniAniSar00}.} \\
\end{tabular}
%\end{ruledtabular}
\end{table}

\subsection{Light unflavored ($u\bar d$) mesons}

The S wave $\pi^+$ and $\rho^+$ spectra are given in
Table~\ref{table:ud}. Masses of both ground states and the first
excited triplet state are used to determine the model parameters
$m_{u/d}$, $c$ and $f$. There is no confirmed datum for the second
excited $\rho^+$ ($3^3$S$_1$). The model prediction is larger than
the experimental value for $4^3$S$_1$ by about 140 MeV. The S wave
$\pi$ spectrum is reproduced by this model very well. The
discrepancies for both $2^1$S$_0$ and $3^1$S$_0$ are within 50
MeV. In particular, this model reproduces the amazingly large mass
difference between $\pi$ and its first excited state.

\begin{table}
\caption{\label{table:us} S wave spectra for strange mesons.
Masses in GeV.}
%\begin{ruledtabular}
\begin{tabular}{||l|l|l@{$\;\;\;\;\;\;\;\;$}||l|l|l@{$\;\;\;\;\;\;\;\;$}||}
 \hline\hline
 $n$ & Experiment$^1$ & Theory    & $n$ & Experiment$^1$& Theory    \\
 \hline\hline
 \multicolumn{3}{||c||}{$^1\mathrm{S}_0$ Singlets $K^+$} &
 \multicolumn{3}{  c||}{$^3\mathrm{S}_1$ Triplets $K^{*+}$} \\
 \hline
 1   & 0.493677(16) & 0.6048    & 1 & 0.89166(26)   & 0.8917    \\
     &              & 0.494$^2$ &   &               & 0.892$^2$ \\
     &              & 0.47$^3$  &   &               & 0.90$^3$  \\
 2   & 1.460$^a$    & 1.5480    & 2 & 1.629(27)$^b$ & 1.6808    \\
     &              & 1.426$^2$ &   &               & 1.649$^2$ \\
     &              & 1.45$^3$  &   &               & 1.58$^3$  \\
 3   & 1.830$^a$    & 2.1040    & 3 & ---           & 2.6242    \\
     &              & 1.976$^2$ &   &               & 2.154$^2$ \\
     &              & 2.02$^3$  &   &               & 2.11$^3$  \\
 \hline\hline
 \multicolumn{6}{l}{$^1$Hagiwara \textit{et al}~\cite{RPP02}; %} \\
                    $^2$Frederico, Pauli and Zhou~\cite{FrePauZho02b};} \\
 \multicolumn{6}{l}{$^3$Godfrey and Isgur~\cite{GodIsg85}. } \\
 \multicolumn{6}{l}{$^a$To be confirmed; $^b$$J^P$ not confirmed.} \\
\end{tabular}
%\end{ruledtabular}
\end{table}

\subsection{Strange ($u\bar s$) mesons}

The S wave $K^+$ and $K^{*+}$ spectra are given in
Table~\ref{table:us}. The mass of the ground state of $K^{*+}$ is
used to determine the mass parameter $m_{s}$. There are many
ambiguities concerning the quantum number assignment for $K$ and
$K^*$ mesons except for the ground states. The model prediction is
larger than the experimental value for the ground state of $K$
($1^1$S$_0$) by about 100 MeV. Note that in \cite{FrePauZho02b},
the mass of $1^1$S$_0$ was used to determine the renormalization
parameter. Both the first and the second excited state of $K$
($2^1$S$_0$ and $3^1$S$_0$) are not confirmed. Another unconfirmed
resonance with mass $1.629\pm 0.027$ GeV lying between $2^1$S$_0$
and $3^1$S$_0$ was assigned to be a singlet $K$. Apparently there
is no position for it in the $K$ spectrum if it is an S wave
state. However, according to its mass, it might well be the first
excited state of $K^*$ ($2^1$S$_0$), according to our model.

\begin{table}
\caption{\label{table:heavy} S wave spectra for heavy mesons.
Masses in GeV.}
\begin{tabular}{||l|l|l@{$\;\;\;\;\;\;\;\;$}||l|l|l@{$\;\;\;\;\;\;\;\;$}||}
 \hline\hline
 $n$ & Experiment$^1$ & Theory    & $n$ & Experiment$^1$& Theory    \\
 \hline\hline
 \multicolumn{3}{||c||}{$^1\mathrm{S}_0$ Singlets $\bar D^0$} &
 \multicolumn{3}{  c||}{$^3\mathrm{S}_1$ Triplets $\bar D^{*0}$} \\
 \hline
 1   & 1.8645(5) & 1.9224    & 1 & 2.0067(5)   & 2.0067    \\
     &           & 1.869$^2$ &   &             & 2.042$^2$ \\
     &           & 1.88$^3$  &   &             & 2.04$^3$  \\
 \hline\hline
 \multicolumn{3}{||c||}{$^1\mathrm{S}_0$ Singlets $B^+$} &
 \multicolumn{3}{  c||}{$^3\mathrm{S}_1$ Triplets $B^{*+}$} \\
 \hline
 1   & 5.2790(5) & 5.2965    & 1 & 5.3250(6)   & 5.3250    \\
     &           & 5.279$^2$ &   &             & 5.325$^2$ \\
     &           & 5.31$^3$  &   &             & 5.37$^3$  \\
 \hline\hline
 \multicolumn{3}{||c||}{$^1\mathrm{S}_0$ Singlets $D^-_s$} &
 \multicolumn{3}{  c||}{$^3\mathrm{S}_1$ Triplets $D^{*-}_s$} \\
 \hline
 1   & 1.9685(6) & 2.0201    & 1 & 2.1124(7)   & 2.0655    \\
     &           & 1.969$^2$ &   &             & 2.069$^2$ \\
     &           & 1.98$^3$  &   &             & 2.13$^3$  \\
 \hline\hline
 \multicolumn{3}{||c||}{$^1\mathrm{S}_0$ Singlets $B^0_s$} &
 \multicolumn{3}{  c||}{$^3\mathrm{S}_1$ Triplets $B^{*0}_s$} \\
 \hline
 1   & 5.3696(24) & 5.3739    & 1 & 5.4166(35)  & 5.3885    \\
     &            & ---       &   &             & 5.3422$^2$ \\
     &            & 5.35$^3$  &   &             & 5.45$^3$  \\
 \hline\hline
 \multicolumn{3}{||c||}{$^1\mathrm{S}_0$ Singlets $B^+_c$} &
 \multicolumn{3}{  c||}{$^3\mathrm{S}_1$ Triplets $B^{*+}_c$} \\
 \hline
 1   & 6.4(4) & 6.4281   & 1 & --- & 6.4327     \\
     &        & ---      &   &     & 6.3458$^2$ \\
     &        & 6.27$^3$ &   &     & 6.34$^3$   \\
 \hline\hline
 \multicolumn{6}{l}{$^1$Hagiwara \textit{et al}~\cite{RPP02}; %} \\
                    $^2$Frederico, Pauli and Zhou~\cite{FrePauZho02b};} \\
 \multicolumn{6}{l}{$^3$Godfrey and Isgur~\cite{GodIsg85}.} \\
\end{tabular}
\end{table}

\subsection{Heavy mesons}

The S wave $u\bar c$, $u\bar b$, $s\bar c$, $s\bar b$ and $c\bar
b$ meson spectra are given in Table~\ref{table:heavy}. No
excitations were observed for these mesons. The present model
agrees with the experiment for heavy mesons better than for light
ones.

The mass of the ground state of $\bar D^{*0}$ is used to determine
the mass parameter $m_{c}$. No much data are available for $D$ and
$D^*$ mesons. The model prediction for $1^1$S$_0$ of $\bar D^{0}$
is very close to the experimental value, deviating from the
experiment by about only 60 MeV.

The mass of the ground state of $\bar B^{*+}$ is used to determine
the mass parameter $m_{b}$. For the ground state of $\bar B^{+}$,
the present model gives a very good agreement with the data.

No experimental values in the $s\bar c$ mesons are used to
determine the model parameters. The model agrees with the
available data of both ground states very well.

No data in $s\bar b$ mesons are used to determine the model
parameters. The model agrees with the experiment very well. The
pseudoscalar spectrum was not calculated for $s\bar b$ mesons in
\cite{FrePauZho02b} because the experimental mass of $1^1$S$_0$ is
larger than the theoretical value of $1^3$S$_1$ thus the
renormalization procedure of \cite{FrePauZho02b} could not be
realized consistently.

The mass of the ground state of $B^+_c$ ($1^1$S$_0$) caries a
large experimental error. The model prediction is within the
experimental limit of errors.

\section{Summary}
\label{sec:summary}

Inspired by \cite{FrePauZho02a,FrePauZho02b} and
\cite{AniAniSar00}, we use a light cone harmonic oscillator model
to study S wave meson spectra, namely the pseudoscalar and vector
mesons. The model Hamiltonian is a mass squared operator
consisting of a quadratic confinement and a hyperfine interaction.
Different from \cite{FrePauZho02a,FrePauZho02b}, the hyperfine
interaction which is responsible for the splitting in the
pseudoscalar-vector spectra acts on both the singlet and triplet
states. The hyperfine interaction is corrected with an universal
fudge factor.

With 4 parameters for the masses of up/down, strange, charm and
bottom quarks, 2 for the harmonic oscillator potential and 1 for
the hyperfine interaction, the model presents a reasonably good
agreement with the 21 available data points.

Finally we remark in all modesty that we do not know any other
model from the literature which reproduces all know S wave mesons
from the lightest ($\pi$) to the heaviest ($B_c$), within the same
model and the same few parameters. Right or wrong, this model at
the worst is very useful for the experimentalists when planning an
experiment. It's simple, transparent and analytical.

\begin{acknowledgments}
S.G.Z. was partly supported by the Major State Basic Research
Development Program of China Under Contract Number G2000077407 and
the National Natural Science Foundation of China under Grant No.
10025522, 10221003 and 10047001.
\end{acknowledgments}


\begin{thebibliography}{99}

\bibitem{BroPauPin98}
 S.~J.~Brodsky, H.~C.~Pauli and S.~S.~Pinsky,
 Phys. Lett. C (Physics Reports) \textbf{301} (1998) 299-486.

\bibitem{FrePauZho02a} T. Frederico, H.~C.~Pauli, and S.~G.~Zhou,
 Phys. Rev. D 66, 054007 (2002).

\bibitem{FrePauZho02b}
 T.~Frederico, H.~C.~Pauli, and S.~G.~Zhou,
 Phys. Rev. D \textbf{66} (2002) 116011
 [arXiv:hep-ph/0210234].

\bibitem{ZhoPau03} S.~G.~Zhou and H.~C.~Pauli,
 \textit{$D^*_{sJ}$(2317)$^+$: a $P$ state from the light cone
         harmonic oscillator model?}
 arXiv: hep-ph/0310330.

\bibitem{Zho03} S.~G.~Zhou,
 \textit{Meson spectra from an effective light cone qcd-inspired model},
 arXiv: hep-ph/0310362.

\bibitem{BjoDre64} Bj\o rken and Drell,
 \textit{Relativistic Quantum Mechanics,} pp.57, 58
 (McGraw-Hill, New York, 1964).

\bibitem{AniAniSar00} A.~V.~Anisovich, V.~V.~Anisovich, and A.~V.~Sarantsev,
 Phys. Rev. D \textbf{62}, 051502(R) (2000).

\bibitem{Pau00b} H.~C.~Pauli,
 Nucl. Phys. \textbf{B} (Proc. Supp.) \textbf{90}, 259 (2000).

\bibitem{Pau03a} H.~C.~Pauli,
 \textit{Succesful renormalization of a QCD-inspired Hamiltonian},
 arXiv: hep-ph/0310294.

\bibitem{RPP02}
 K.~Hagiwara~\textit{et al.},
 Phys. Rev. D \textbf {66} (2002) 010001.

\bibitem{GodIsg85}
 S.~Godfrey and N.~Isgur,
 Phys. Rev. D \textbf{32} (1985) 189.

\end{thebibliography}
\end{document}